\documentclass[fleqn,12pt]{article}
\usepackage{epsfig}
\begin{document}
\textheight 22cm
\textwidth 15cm
\noindent
{\Large \bf The momentum flux probability distribution function for ion-temperature-gradient turbulence}
\newline
\newline
Johan Anderson\footnote{anderson.johan@gmail.com} and Eun-jin Kim
\newline
University of Sheffield
\newline
Department of Applied Mathematics
\newline
Hicks Building, Hounsfield Road 
\newline 
Sheffield
\newline
S3 7RH
\newline
UK
\newline
\newline
\begin{abstract}
\noindent
There has been overwhelming evidence that coherent structures play a critical role in determining the overall transport in a variety of systems. We compute the probability distribution function (PDF) tails of momentum flux and heat flux in ion-temperature-gradient turbulence, by taking into account the interaction among modons, which are assumed to be coherent structures responsible for bursty and intermittent events, contributing to the PDF tails. The tail of PDF of momentum flux $R = \langle v_x v_y \rangle$ is shown to be exponential with the form $\exp{\{-\xi R^{3/2}\}}$, which is broader than a Gaussian, similarly to what was found in the previous local studies. An analogous expression with the same functional dependence is found for the PDF tails of heat flux. Furthermore, we present a detailed numerical study of the dependence of the PDF tail on the temperature and density scale lengths and other physical parameters through the coefficient $\xi$.
\end{abstract}
\newpage
\renewcommand{\thesection}{\Roman{section}}
\section{Introduction}
One of the main challenges in magnetic fusion research has been to predict the turbulent heat and particle transport originating from various micro-instabilities. The ion-temperature-gradient (ITG) mode is one of the main candidates for causing the anomalous heat transport in core plasmas of tokamaks~\cite{a10}. Significant heat transport might however be mediated by coherent structures such as streamers and blobs through the formation of avalanche like events of large amplitude~\cite{a14}-~\cite{a16}, as indicated by recent numerical studies. These events cause the deviation of the probability distribution functions (PDFs) from a Gaussian profile on which the traditional mean field theory (such as transport coefficients) is based. In particular, PDF tails due to rare events of large amplitude are often found to be substantially different from Gaussian although PDF centers tend to be Gaussian~\cite{a38}. These non-Gaussian PDF tails are manifestations of intermittency, caused by bursts and coherent structures. The characterization of these PDF tails thus requires a non-perturbative method.

There are many coherent structures of interest, including zonal flows and streamers. While streamer-like structures ($k_{\theta} \gg k_{r}$) enhance transport zonal flows can dramatically reduce transport~\cite{a35}-~\cite{a36}. The zonal flows are poloidally and toroidally symmetric ($k_{\theta} = 0$, $k_{\parallel} = 0$) and radially inhomogeneous ($k_r \neq 0$) flow structures in toroidal plasmas. It would thus be of great importance to develop a theory of the formation of coherent structures and PDFs of heat flux due to these structures.

The purpose of this paper is to investigate the likelihood of the formation of coherent structures by computing the PDF (tails) of the Reynolds stress and predict the PDFs of heat transport. Specifically, we extend a non-perturbative theory of the PDFs of local momentum flux and heat flux to accommodate global fluxes by incorporating the interactions among structures. An advanced fluid model for the ITG mode is used~\cite{a13} that has been successful in reproducing both experimental~\cite{a11} and non-linear gyro-kinetic results~\cite{a12}. The advanced fluid model is expected to give qualitatively and quantitatively accurate results in tokamak core plasmas or the flat density regime. For instance, this was shown in the plot of the slope of the ITG threshold as a function of $\epsilon_n$ in the flat density regime (see Figure 1 in Ref.~\cite{a44}) where this model recovers the full linear kinetic result within 5\%, far better than the prediction from an approximate kinetic model using a constant energy approximation in the magnetic drift. In particular, the gyrokinetic simulations of ITG mode turbulence reported in the Cyclone work Ref.~\cite{a12} indicate that there is a strong excitation of zonal flows close to marginal stability where the non-linearly generated flows were able to damp out the turbulence resulting in a non-linear up-shift in the critical temperature gradient needed to obtain transport for longer time scales. One particular coherent structure in ITG drift wave turbulence is the two dimensional bipolar vortex soliton solution called the modon~\cite{a25}-~\cite{a27}. The modon is an exact solution to the non-linear governing equations for ITG mode turbulence and travels perpendicular to both the strong magnetic field and the background density gradient. The PDF tail is viewed as the transition amplitude from an initial state with no fluid motion to a final state governed by the modon with different amplitudes in the long time limit.  

The theoretical technique used here is the so-called instanton method, a non-perturbative way of calculating the PDF tails. The PDF tail is first formally expressed in terms of a path integral by utilizing the Gaussian statistics of the forcing. An optimum path will then be associated with the creation of a modon (among all possible paths) and the action is evaluated using the saddle-point method on the effective action. The saddle-point solution of the dynamical variable $\phi(x,t)$ of the form $\phi(x,t) = F(t) \psi(x)$ is called an instanton if $F(t) = 0$ at the initial time and $F(t) \neq 0$ in the long time limit. The instanton is localized in time, existing during the formation of the modon. Thus, the bursty event can be associated with the creation of a modon. Note that, the function $\psi(x)$ here represents the spatial form of the coherent structure. Historically, the instanton method was used in gauge field theory for calculating the transition amplitude from one vacuum to another vacuum~\cite{a24}-~\cite{a241}.

We will show exponential PDF tails for both heat flux ($H$) and momentum fluxes ($R$) of the form $e^{-\xi H^{3/2}}$ and $e^{-\xi R^{3/2}}$, similar to the PDFs of local fluxes in the earlier works~\cite{a19}-~\cite{a21}. The dependence of the coefficient $\xi$ (i.e. the overall amplitude of the PDF tails) on parameter values (e.g. the density gradient, temperature gradient, curvature, etc) will be studied in detail. We will also demonstrate the generation of monopole vortex via the interaction of two dipoles (modons).

The paper is organized as follows. In Sec. II the model for the ITG mode is presented together with preliminaries for the path-integral formulation for the PDF tail of momentum flux. In Sec. III the instanton solutions are calculated and the PDF tail of momentum flux is estimated in Sec IV. We provide numerical results in Sec. V and a discussion of the results and the PDF tail of heat flux and conclusion in Sec. VI.

\section{Non-perturbative calculation of momentum flux PDF}
The ITG mode turbulence is modeled using the continuity and temperature equation for the ions and considering the electrons to be Boltzmann distributed; quasi-neutrality is used to close the system~\cite{a13}. In the present work the effects of parallel ion motion, magnetic shear, trapped particles and finite beta on the ITG modes are neglected since in previous works that the effect of parallel ion motion on the ITG mode was found to be rather weak~\cite{a37}.

The continuity and temperature equations are,
\begin{eqnarray}
\frac{\partial \tilde{n}}{\partial t} - \left(\frac{\partial}{\partial t} - \alpha_i \frac{\partial}{\partial y}\right)\nabla^2_{\perp} \tilde{\phi} + \frac{\partial \tilde{\phi}}{\partial y} - \epsilon_n g_i \frac{\partial}{\partial y} \left(\tilde{\phi} + \tau \left(\tilde{n} + \tilde{T}_i \right) \right) + \nu \nabla^4 \tilde{\phi} = \nonumber \\
- \left[\tilde{\phi},\tilde{n} \right] + \left[\tilde{\phi}, \nabla^2_{\perp} \tilde{\phi} \right] + \tau \left[\tilde{\phi}, \nabla^2_{\perp} \left( \tilde{n} + \tilde{T}_i\right) \right] + f,\\
\frac{\partial \tilde{T}_i}{\partial t} - \frac{5}{3} \tau \epsilon_n g_i \frac{\partial \tilde{T}_i}{\partial y} + \left( \eta_i - \frac{2}{3}\right)\frac{\partial \tilde{\phi}}{\partial y} - \frac{2}{3} \frac{\partial \tilde{n}}{\partial t} = \nonumber \\
- \left[\tilde{\phi},\tilde{T}_i \right] + \frac{2}{3} \left[\tilde{\phi},\tilde{n} \right].
\end{eqnarray}
Here $\left[ A ,B \right] = (\partial A/\partial x) (\partial B/\partial y) - (\partial A/\partial y) (\partial B/\partial x)$ is the Poisson bracket; $f$ is a forcing; $\tilde{n} = (L_n/\rho_s) \delta n / n_0$, $\tilde{\phi} = (L_n/\rho_s) e \delta \phi /T_e$, $\tilde{T}_i = (L_n/\rho_s) \delta T_i / T_{i0}$ are the normalized ion particle density, the electrostatic potential and the ion temperature, respectively. In equations (1) and (2), $\tau = T_i/T_e$, $\rho_s = c_s/\Omega_{ci}$ where $c_s=\sqrt{T_e/m_i}$, $\Omega_{ci} = eB/m_i c$ and $\nu$ is collisionality. We also define $L_f = - \left( d ln f / dr\right)^{-1}$ ($f = \{n, T_i \}$), $\eta_i = L_n / L_{T_i}$, $\epsilon_n = 2 L_n / R$ where $R$ is the major radius and $\alpha_i = \tau \left( 1 + \eta_i\right)$. The perpendicular length scale and time are normalized by $\rho_s$ and $L_n/c_s$, respectively. The geometrical quantities are calculated in the strong ballooning limit ($\theta = 0 $, $g_i \left(\theta = 0, \kappa \right) = 1/\kappa$  where $g_i \left( \theta \right)$ is defined by $\omega_D \left( \theta \right) = \omega_{\star} \epsilon_n g_i \left(\theta \right)$)~\cite{a28}-~\cite{a29}, with $\omega_{\star} = k_y v_{\star} = \rho_s c_s k_y/L_n $. The system is closed by using quasi-neutrality with Boltzmann distributed electrons. Note that the approximate linear solutions to Equations (1) and (2) with $f = 0$ give the dispersion relation with real frequency ($\omega_r$) and growth rate ($\gamma$) as,
\begin{eqnarray}
\omega_{r} & = & \frac{k_y}{2\left( 1 + k_{\perp}^2\right)} \left( 1 - \left(1 + \frac{10\tau}{3} \right) \epsilon_n g_i - k_{\perp}^2 \left( \alpha_i + \frac{5}{3} \tau \epsilon_n g_i \right)\right),  \\
\gamma & = & \frac{k_y}{1 + k_{\perp}^2} \sqrt{\tau \epsilon_n g_i \left( \eta_i - \eta_{i th}\right)},
\end{eqnarray}
where 
\begin{eqnarray}
\eta_{i th} \approx \frac{2}{3} - \frac{1}{2 \tau} + \frac{1}{4 \tau \epsilon_n g_i} + \epsilon_n g_i \left( \frac{1}{4 \tau} + \frac{10}{9 \tau}\right).
\end{eqnarray}
Finite Larmor Radius (FLR) effects on the $\eta_{i th}$ are here neglected, although these are included in the numerical study leading to Figures 1 and 2.

We formally calculate the PDF tails of momentum flux by using the instanton method. To this end, the PDF tail is expressed in terms of a path integral by utilizing the Gaussian statistics of the forcing $f$~\cite{a22}. The probability distribution function for Reynolds stress $R$ can be defined as
\begin{eqnarray}
P(R) & = &  \langle \delta(\langle v_x v_y \rangle - R) \rangle \nonumber \\
& = & \int d \lambda  \exp(i \lambda R) \langle \exp(-i \lambda (v_x v_y)) \rangle \nonumber \\
& = & \int d\lambda \exp(i \lambda R) I_{\lambda},
\end{eqnarray}
where 
\begin{eqnarray}
I_{\lambda} = \langle \exp(-i \lambda v_x v_y) \rangle.
\end{eqnarray}
The integrand can then be rewritten in the form of a path-integral as
\begin{eqnarray}
I_{\lambda} = \int \mathcal{D} \phi \mathcal{D} \bar{\phi} e^{-S_{\lambda}}.
\end{eqnarray}
The angular brackets denote the average over the statistics of the forcing $f$. By using the ansatz $T_i = \chi \phi$, which will be justified later (see Eq. (15)), the effective action $S_{\lambda}$ in Eq. (8) can be expressed as
\begin{eqnarray}
S_{\lambda} & = & -i \int d^2x dt \bar{\phi} \left( \frac{\partial \phi}{\partial t} - (\frac{\partial }{\partial t} - \alpha_i \frac{\partial }{\partial y}) \nabla^2_{\perp} \phi + (1-\epsilon_n g_i \beta)\frac{\partial \phi}{\partial y} - \beta [\phi, \nabla^2_{\perp} \phi ]\right) \nonumber \\
& + & \frac{1}{2} \int dt d^2x d^2 x^{\prime} \bar{\phi}(x) \kappa(x-x^{\prime}) \bar{\phi}(x^{\prime}) \nonumber \\
& + & i \lambda \int d^2 x dt (-\frac{\partial \phi}{\partial x} \frac{\partial \phi}{\partial y}) \delta(t).
\end{eqnarray}
Here, 
\begin{eqnarray}
\beta & = & 1 + \tau + \tau \chi, \\
\chi & = & \frac{\eta_i - \frac{2}{3}(1-U)}{U+\frac{5}{3}\tau \epsilon_n g_i}.
\end{eqnarray}

In Eq. (11), $U$ is the modon speed (see Eq. (15)).
To obtain Eq. (9) we have assumed the statistics of the forcing $f$ to be Gaussian with a short correlation time modeled by the delta function as
\begin{eqnarray}
\langle f(x, t) f(x^{\prime}, t^{\prime}) \rangle = \delta(t-t^{\prime})\kappa(x-x^{\prime}),
\end{eqnarray}
and $\langle f \rangle = 0$.
The delta correlation in time were chosen for the simplicity of the analysis. In the case of a finite correlation time the non-local integral equations in time are needed. We will also make use of the completeness of the Bessel function expansion and write $\kappa(x-x^{\prime}) = \kappa_0 (J_0(kx)J_0(kx^{\prime})+J_1(kx)J_1(kx^{\prime})(\cos \theta \cos \theta^{\prime} + \sin \theta \sin \theta^{\prime})+J_2(kx)J_2(kx^{\prime})(\cos 2 \theta \cos 2 \theta^{\prime} + \sin 2 \theta \sin 2 \theta^{\prime}) + ...)$.
\section{Instanton (saddle-point) solutions}
We have now reformulated the problem of calculating the PDF to a path-integral as in Eq. (6). Although the path integral cannot in general be calculated exactly, an approximate value can be found in the limit $\lambda \rightarrow \infty$ by using a saddle point method. The idea of the saddle-point method is that the integrand has a unique global maximum and that all significant contributions to the integral  come only from points in the vicinity of this maximum. In the limit $\lambda \rightarrow \infty$ a particular path that satisfies the saddle-point equations gives the leading order contribution. The saddle-point equations are;
\begin{eqnarray}
\frac{\delta S_{\lambda}}{\delta \phi} & = & 0, \\
\frac{\delta S_{\lambda}}{\delta \bar{\phi}} & = & 0.
\end{eqnarray} 
Since a direct application of the saddle-point equations results in very complicated partial differential equations for $\phi$ and $\bar{\phi}$, we assume that the instanton saddle-point solution is a temporally localized modon. That is, we assume that a non-linear vortex soliton solution exists to the system of Eqs (1)-(2) by assuming that the electric potential $\phi$ can be written
\begin{eqnarray}
\phi(x,y,t) = \psi(x,y-Ut)F(t), \mbox{and   } T_i = \chi \phi.
\end{eqnarray}
The function $\psi$ in Eqs (16)-(17) is the spatial form of the coherent structure, which is assumed to be the sum of the two modons with the ratio of strength $\epsilon$ as follows, 
\begin{eqnarray}
\psi(x,y-Ut)& = & c_1 J_1(kr) (\cos \theta + \epsilon \sin \theta ) + \frac{\alpha}{k^2} r \cos \theta \mbox{   for   } r \leq a, \\
\psi(x,y-Ut)& = & c_2 K_1(pr) (\cos \theta + \bar{\epsilon}(r) \sin \theta )   \mbox{   for   } r \geq a. 
\end{eqnarray}
Here $J_1$ and $K_1$ are the first Bessel function and the second modified Bessel function, respectively. Here $r = \sqrt{x^2+y^2}$, $\tan \theta = y^{\prime}/x$, $y^{\prime} = y - Ut$, $\alpha = (A_1 - k^2 A_2)$, $A_1 = (1-\epsilon_n g_i - U)/\beta$, $A_2 = (U+\alpha_i)/\beta$. By matching the inner and outer solution at $r=a$ we find the conditions $c_1 = -\alpha a/J_1(ka)$, $c_2 = -Ua/K_1(pa)$, $J_1^{\prime}(ka)/J_1(ka) = (1+k^2/p^2)/ka - kK_1^{\prime}(pa)/pK_1(pa)$; $U$ is the velocity of the modon, and $a$ is the size of the core region. The function $\bar{\epsilon}(r)$ is chosen such that the matching conditions are similar to those in previous previous studies~\cite{a19}-~\cite{a20}. 

It is important to note that if $\epsilon = 0$ that the averaged Reynolds stress vanishes. In general the Reynolds stress associated with the modon can be found to be
\begin{eqnarray}
R_0 & = & \langle v_x v_y \rangle  =  \int d^2x (-\frac{\partial \phi}{\partial x} \frac{\partial \phi}{\partial y}) \nonumber \\
& = &- \epsilon \pi c_1^2 \int dr r [\frac{\alpha}{k c_1} J_0(kr) + \frac{1}{4} (\frac{J_1(kr)}{r})^2 + \frac{3}{2} k J_1^{\prime}(kr) \frac{J_1(kr)}{r} \nonumber \\
& + & \frac{1}{4}k^2 (J_1^{\prime}(kr))^2 + \frac{1}{4} k^2 (J_2(kr))^2].
\end{eqnarray}
The Reynolds stress in Eq. (18) represents the effective force driving coherent structures. Interestingly, it shows that the non-linear interaction of two modons (dipoles) can generate a monopole, given by the zeroth order Bessel function ($J_0$), as well as other more complicated structures. Given that a monopole is the dominant term our result, Eq. (18), indicates the tendency of monopole formation from modons. This tendency was observed in numerical studies in Ref.~\cite{a27},~\cite{a31}. Next, the action $S_{\lambda}$ is to be expressed only as an integral in time by using the conjugate variables
\begin{eqnarray}
\bar{F}_0 & = & \int d^2 x \bar{\phi}(x,t) J_0(kr), \\
\bar{F}_{1s} & = & \int d^2 x \bar{\phi}(x,t) J_1(kr) \sin \theta, \\
\bar{F}_{1c} & = & \int d^2 x \bar{\phi}(x,t) J_1(kr) \cos \theta, \\
\bar{F}_{2s} & = & \int d^2 x \bar{\phi}(x,t) J_2(kr) \sin 2 \theta, \\
\bar{F}_{2c} & = & \int d^2 x \bar{\phi}(x,t) J_2(kr) \cos 2 \theta.
\end{eqnarray}
Note that the contribution from the outer solution ($r>a$) to $S_{\lambda}$ is neglected compared to that from the inner solution ($r<a$) for simplicity. The outer solution decays fast and inherently gives a minor contribution to the PDF tail. The action $S_{\lambda}$ consists of three different parts; the ITG model, the forcing and the Reynolds-stress parts respectively. The ITG model part of the action can be reduced to
\begin{eqnarray}
I_{ITG} & = & -i \int d^2x dt \left[ \dot{F}((1+k^2)\psi - k^2 \alpha x) \right. \nonumber \\
& + & \left. F\left[1  - \epsilon_n g_i \beta - U - k^2(U+\alpha_i) + \beta F k^2 \alpha \right] \frac{\partial \psi}{\partial y} \right. \nonumber \\
& + & \left. F \nu k^4 \psi - F \nu k^4 \alpha x \right].
\end{eqnarray}

The full action including the forcing and Reynolds stress terms can then be expressed in terms of $F$, $\dot{F}$ and the conjugate variables $\bar{F}$
\begin{eqnarray}
S_{\lambda} & = & -i \int dt [\gamma_1 \dot{F}(\bar{F}_{1c} + \epsilon \bar{F}_{1s})+  F(\gamma_2 \bar{F}_{2s} + \epsilon \gamma_3 \bar{F}_{0} + \epsilon \gamma_4 \bar{F}_{2c}) \nonumber \\
& + & F^2(\gamma_5 \bar{F}_{2s} + \epsilon \gamma_6 \bar{F}_{0} + \epsilon \gamma_7\bar{F}_{2c}) +  \gamma_8 (\bar{F}_{1c} + \epsilon \bar{F}_{1s}) )] \nonumber \\
& + & \frac{1}{2} \kappa_0 \int dt (\bar{F}_{0}^2 + 2(\bar{F}_{1c}^2 + \bar{F}_{1s}^2) + 2(\bar{F}_{2s}^2 + \bar{F}_{2c}^2)) \nonumber \\
& - & \lambda R_0 \int dt F^2 \delta(t)
\end{eqnarray}
Here the coefficients are;
\begin{eqnarray}
\gamma_1 & = & c_1(1 + k^2 + \frac{2 \alpha}{k^3}), \\
\gamma_2 & = & -\frac{k}{2} \alpha_1 = - \gamma_3 = - \frac{1}{2} \gamma_4, \\
\gamma_5 & = & -\frac{k}{2} \beta \alpha = - \gamma_6, \\
\gamma_7 & = & \beta k \alpha, \\
\gamma_8 & = & \nu k^4, \\
\alpha_1 & = & 1 - \epsilon_n g_i\beta - U - k^2(U+\alpha_i).
\end{eqnarray}

The equations of motions for the instanton are found by the variations of the action with respect to $F$, $\bar{F}_{0}$, $\bar{F}_{1c}$, $\bar{F}_{1s}$, $\bar{F}_{2c}$ and $\bar{F}_{2s}$;
\begin{eqnarray}
\frac{\delta S_{\lambda}}{\delta F} & = & -i[-\gamma_1 (\dot{\bar{F}}_{1c} + \epsilon \dot{\bar{F}}_{1s}) + (\gamma_2 \bar{F}_{2s} + \epsilon \gamma_3 \bar{F}_{0} + \epsilon \gamma_4 \bar{F}_{2c}) \nonumber \\
& + & 2F (\gamma_5 \bar{F}_{2s} + \epsilon \gamma_6 \bar{F}_{0} + \epsilon \gamma_7\bar{F}_{2c}) +  \gamma_8 (\bar{F}_{1c} + \epsilon \bar{F}_{1s}) ] \nonumber \\
& - & 2 \lambda R_0 F \delta(t) = 0, \\
\frac{\delta S_{\lambda}}{\delta \bar{F}_0} & = & -i\epsilon ( \gamma_3 F + \gamma_6 F^2) + \kappa_0 \bar{F}_0 = 0, \\
\frac{\delta S_{\lambda}}{\delta \bar{F}_{1c}} & = & -i(\gamma_1 \dot{F} + \gamma_8 F) + 2 \kappa_0 \bar{F}_{1c} = 0, \\
\frac{\delta S_{\lambda}}{\delta \bar{F}_{1s}} & = & -i\epsilon (\dot{F} + \gamma_8 F) + 2 \kappa_0 \bar{F}_{1s} = 0, \\
\frac{\delta S_{\lambda}}{\delta \bar{F}_{2c}} & = & -i\epsilon (\gamma_4 F + \gamma_7 F^2) + 2 \kappa_0 \bar{F}_{2c} = 0, \\
\frac{\delta S_{\lambda}}{\delta \bar{F}_{2s}} & = & -i (\gamma_2 F + \gamma_5 F^2) + 2 \kappa_0 \bar{F}_{2s} = 0.
\end{eqnarray}
The equation of motion for $F$ is derived for $t < 0$ using Eqs. (32)-(37) as;
\begin{eqnarray}
\frac{1}{2} \gamma_1^2(1 + \epsilon^2 ) \frac{d \dot{F}^2}{dF}& = & \eta_1 F + 3 \eta_2 F^2 + 2 \eta_3 F^3, \\
\eta_1 & = & \gamma_2^2 + 2 \epsilon^2 \gamma_3^2 + \epsilon^2 \gamma_4^2 + \gamma_8^2 + \epsilon^2 \gamma_8^2, \\
\eta_2 & = & \gamma_2 \gamma_5 + 2 \epsilon^2 \gamma_3 \gamma_6 + \epsilon^2 \gamma_4 \gamma_7, \\
\eta_3 & = & \gamma_5^2 + 2 \epsilon^2 \gamma_6^2 + \epsilon^2 \gamma_7^2. 
\end{eqnarray}
The contribution from the dissipation ($\nu$) to the term involving the time derivative of $F$ ($\dot{F}$) cancels out and the equation of motion is exactly solvable. In the limit of $\lambda \rightarrow \infty$ the relation can be written,
\begin{eqnarray}
\dot{F} \simeq \sqrt{\frac{\eta_3}{\gamma_1^2(1 + \epsilon^2)}} F^2.
\end{eqnarray}
The initial condition is found by integrating Eq. (32) over the interval $[-\delta, 0]$ ($\delta \ll 1 $) and observing that the conjugate variables mediating between the forcing and $F$ vanish for $t \geq 0$. This can be interpreted as a ``causality'' condition
\begin{eqnarray}
i\gamma_1 (\bar{F}_{1c}(-\delta) + \epsilon \bar{F}_{1s}(-\delta)) + 2 \lambda R_0 F(0) = 0. 
\end{eqnarray}
The elimination of $\bar{F}_{1c}$ and $\bar{F}_{1s}$ using Eqs. (34) and (35) gives a relationship between $\dot{F}$ and $F$ in the limit $\lambda \rightarrow \infty $ as,
\begin{eqnarray}
\gamma_1^2(1 + \epsilon^2)\dot{F}(-\delta) \simeq 4 \kappa_0 \lambda R_0 F(0).
\end{eqnarray}
Finally, Eqs. (43) and (44) gives the initial condition for $F$
\begin{eqnarray}
F(0) \simeq \frac{4 \kappa_0 \lambda R_0}{\sqrt{\eta_3 \gamma_1^2(1 + \epsilon^2)}}.
\end{eqnarray}
We use this initial condition to compute the saddle-point action and then predict the scaling of $S_{\lambda}$ (as $\lambda \rightarrow \infty$) to compute the PDF tail in the next section.

\section{The PDF tail}
The PDF tail is found by calculating the value of $S_{\lambda}$ at the saddle-point. Of particular interest is the dependency of the action on $\lambda$ in the limit $\lambda \rightarrow \infty$ which will give the PDF tail;
\begin{eqnarray}
S_{\lambda} & \simeq & - \frac{1}{3} \lambda R_0 \left(\frac{4 \lambda \kappa_0 R_0}{\sqrt{\eta_3 \gamma_1^2(1 + \epsilon^2)}}\right)^2 \nonumber \\ 
& \simeq & - \frac{1}{3}h \lambda^3. 
\end{eqnarray}
The PDF tail of the Reynolds stress ($R$) can now be found by performing the integration over $\lambda$ in Eq. (6) using the saddle-point method,
\begin{eqnarray}
P(R) & \sim & \exp(-\xi (\frac{R}{R_0})^{3/2}), \\
\xi & = &  \frac{2}{3}\frac{\sqrt{\eta_3\gamma_1^2(1 +\epsilon^2)}}{4 \kappa_0}.
\end{eqnarray}
Here, it is important to note that our assumption that $\lambda \rightarrow \infty$ corresponds to $R \rightarrow \infty$. Equation (47) gives the probability of a Reynolds stress $R$, normalized by the Reynolds stress $R_0$ due to the the modon solution $R_0$ given by Eq. (18) which is fixed for given parameters. We have assumed that the modon is created and that $F(t) = 0$ as $t \rightarrow - \infty$, which means we can interpret the Eq. (47) as a transition amplitude from an initial state, with no fluid motion, to final states with different values of $R/R_0$. It is also important to note that even though we have assumed a Gaussian forcing the tail is non-Gaussian exhibiting intermittency. The overall coefficient $\xi$ in the PDF in Eqs. (47)-(48) depends on several physical parameter values. First, $\xi \rightarrow \infty$ (i.e. PDF vanishes) as the the forcing disappears ($\kappa_0 \rightarrow 0$); the instanton cannot form and the PDF vanishes ($P(R) \rightarrow 0$). Second, $\xi$ increases for larger $\epsilon$ leading to a reduction in the PDF. Furthermore, $\xi$ depends on ion temperature gradient ($\eta_i$), density gradient ($\epsilon_n$), temperature ratio ($\tau=T_i/T_e$), modon size ($a$), modon speed ($U$) and wave number ($k$). These dependencies will be studied in detail numerically in Sec. V.
\section{Results}
We have presented a calculation of the tail of the Reynolds stress probability distribution function (PDF) in forced ITG turbulence. We have shown by a non-perturbative calculation (instanton calculus) that a coherent structure can lead to intermittent vorticity flux. A path-integral formulation is developed for the PDF tail. The system is solved by assuming that the tail is associated with bursty events or the creation of modons (a bipolar vortex). The integrals are then estimated by the saddle-point method, which gives the functional dependence of the tail as $\exp{\{-\xi (R/R_0)^{3/2}\}}$. This exponential form seems to be ubiquitous in drift wave turbulence. In this section, the parameter dependencies of $\xi$ will be studied in detail and compared to a normalized Gaussian profile.

In Figure 1, the PDF tails are shown as a function of Reynolds stress for forced ITG mode turbulence (blue line), forced Hasegawa-Mima (HM) turbulence (red line), Gaussian distributions with same parameters as for the ITG mode turbulence (green line) and cases with negative modon speed in ITG mode turbulence (black line) and HM turbulence (dashed black line). The other parameters are $\eta_i = 4.0 $, $\tau = 0.5$, $\epsilon_n = 1.0$, $g_i = 1$, $a = 2$, $U= 2.0$ ($U = -5.0$ black line and dashed black line), $\kappa_0 = 3.0$, $\epsilon = 0.1$ and $k \approx 1.84$ (ITG case), $k=0.81$ (ITG with reversed modon speed), $k \approx 1.73$ (HM case) and $k=1.56$ (HM with reversed modon speed). The PDF tail in HM model is obtained by setting the parameters $\beta = 1$ and $\epsilon_n = 0.0$. This is equivalent to letting $\eta_i = 0.0$ and $\tau=T_i/T_e = 0.0$. It is shown that for $U>0$ the PDF tail for ITG turbulence is significantly lower compared to the value found for HM turbulence, although they are qualitatively similar. The reason for this is that the model dependent factor $\xi$ is significantly different in ITG mode and HM turbulence. In the case of reversed modon speed ($U<0$), $\chi$ changes sign and significantly enhances the PDF tail. The enhanced ITG PDF tail for the reversed modon speed is due to the change in $k$; a smaller value of $k$ combined with a large negative modon speed gives the enhancement.

\begin{figure}
  \includegraphics[height=.3\textheight]{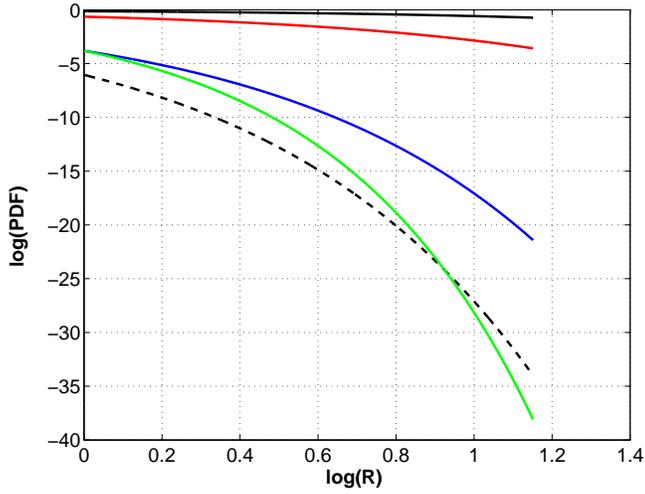}
  \caption{(Color online). The PDF tails are shown as a function of Reynolds stress for ITG mode turbulence (blue line), forced Hasegawa-Mima (HM) turbulence (red line), Gaussian distributions with same parameters as for the ITG mode turbulence (green line) and cases with negative modon speed in ITG mode turbulence (black line) and in HM turbulence (dashed black line).  The other parameters are $\eta_i = 4.0 $, $\tau = 0.5$, $\epsilon_n = 1.0$, $g_i = 1$, $a = 2$, $U= 2.0$ ($U = -5.0$ black line and dashed black line), $\kappa_0 = 3.0$, $\epsilon = 0.1$ and $k \approx 1.84$ (ITG case), $k=0.81$ (ITG with reversed modon speed), $k \approx 1.73$ (HM case) and $k=1.56$ (HM with reversed modon speed).}
\end{figure}

Figure 2 shows the PDF tail as a function of Reynolds stress in ITG mode turbulence by varying $\epsilon_n$ parameter values; $k \approx 1.91$, $\epsilon_n = 0.1$ (black), $\epsilon_n = 1.0$, (blue line), $\epsilon_n = 2.0$, (red line) and a Gaussian distribution with the same $\xi$ as for the $\epsilon_n = 1.0$ case (green line). The parameters are $U=4.0$ with all the others the same as those in Figure 1. The PDF tails for different density profiles are qualitatively similar, however for peaked density profiles (small $\epsilon_n$) the PDF tails are significantly enhanced.

\begin{figure}
  \includegraphics[height=.3\textheight]{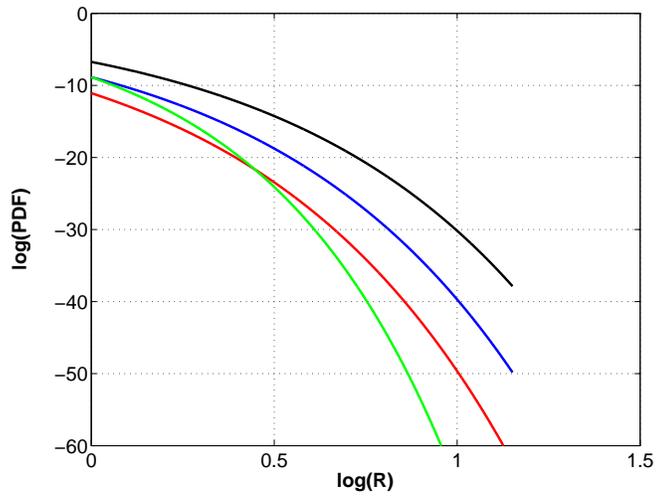}
  \caption{(Color online). Shows the PDF tail as a function of Reynolds stress in ITG mode turbulence by varying $\epsilon_n$ parameter values; $\epsilon_n = 0.1$, $k \approx 1.91$ (black), $\epsilon_n = 1.0$, (blue line), $\epsilon_n = 2.0$, (red line) and a Gaussian control distribution with the same $\xi$ as for the $\epsilon_n = 1.0$ case (green line). The parameters are $U=4.0$ and the others as in Figure 1.}
\end{figure}

In most cases, the PDF tails of drift wave turbulence differ significantly both quantitatively and qualitatively with the Gaussian distributions. 

\section{Discussion and conclusions}
The present calculation of the PDF tail of momentum suggests the PDF tail $\sim e^{-\xi R^{3/2}}$ to be ubiquitous in drift wave turbulence. The 3/2 exponent in Eq. (47) follows from the $\lambda \rightarrow \infty$ dependence of the action as $S_{\lambda} \sim \lambda^3$ and inherently comes from the quadratic non-linearity in the dynamical equations. This scaling can be found by balancing the terms in $S_{\lambda}$ as $\lambda \rightarrow \infty$ as $\lambda \phi \phi \sim \phi \bar{\phi} \sim T \bar{\phi} \phi^2 \sim T \bar{\phi}^2$, from which it follows: $\phi \sim \lambda,\ \bar{\phi} \sim \lambda^2,\ T \sim \lambda^{-1},$ and $S_{\lambda} \sim \lambda^3$. Here, $T$ is the typical timescale of the instanton. The coefficient $\xi$ in Eq. (47) contains all the model dependent information and gives the difference between the ITG and the HM models; we have shown that PDF tails in drift wave turbulence significantly deviate from the Gaussian distributions in most cases. Moreover, it was found that reversed modon speed may have significant influence on the PDF tail (e.g see Figure 1). There is some interesting recent experimental work done at CSDX at UCSD, where the same qualitative Reynolds stress PDF scaling was found ($e^{- \xi R^{3/2}}$)~\cite{a43}, which agrees with our prediction.

We now consider the PDF of ion heat flux $\langle T_i v_x \rangle$ given by,

\begin{eqnarray}
P(H) & \sim &  \langle \delta(\langle T_i v_x \rangle - H) \rangle \nonumber \\
& = & \int d \lambda  \exp(i \lambda H) \langle \exp(-i \lambda (T_i v_x)) \rangle, 
\end{eqnarray}
By using the same methodology, we can show that $P(H)$ takes exactly the same form as for $P(R)$; i.e. $P(H) \sim e^{-\xi (H/H_0)^{3/2}}$. Here, $H_0 = \langle T_i v_x \rangle$, is the heat flux associated with modons. Therefore, heat transport can also be significantly enhanced over the Gaussian prediction due to modons.

In using the instanton method we had to assume the spatial form of the coherent structure, a modon in this case, to be an exact solution to the non-linear dynamical equation with fixed parameter values. This means that the mechanism of formation of the structure itself was not addressed in this analysis. The computation of structure formation will be addressed in a future publication. 

We note that for simplicity, we have not included the effect of zonal flows in the present study. The system of Eq. (1)-(2) describes the non-linear evolution of the ITG mode in the presence of a Gaussian forcing ($f$), by assuming the Boltzmann response. The influence of zonal flows on the PDFs can be calculated by using a similar method, and is expected to give a qualitatively similar result ($e^{-\xi R^{3/2}}$), but with a different value of the coefficient $\xi$. As long as the main non-linearity is quadratic in Eq. (1), the same power law should follow since the power-law is determined by the highest nonlinearity in the equations. Due to the neglect of zonal flows the result obtained in this paper would be more relevant to understanding transport in drift-wave systems that have weak influence of zonal flows e.g. electron-temperature-gradient (ETG) turbulence. However, even if zonal flows play an important role in the regulation of ITG turbulence, it is still of great interest to investigate the PDF tails in ITG turbulence which is weakly regulated by zonal flows. It is because many transport simulations have shown that the zonal flows have strong effect close to the critical gradient whereas far away from this critical gradient the effect of zonal flows is much more complicated.  

In general, for calculating the PDF tail a weighted sum over various coherent structures is needed. At present, the only known exact solution is the modon which we have assumed to be the underlying coherent structure. A generalization should be straight forward if more non-linear solutions were available.

We note that a non-Gaussian scaling of the PDF (the exponent of R) is found even when the forcing is Gaussian, although the exact exponent may depend on the temporal and possibly spatial correlation of the forcing ($f$). In the present paper, the forcing is chosen to be temporally delta correlated for simplicity. The exponent may also change, if another spatial coherent structure is introduced i.e. another non-linear solution to the Eq. (1)-(2) is found. The case where the forcing is non-Gaussian ($f$) will be addressed in a future publication.

Although there are very few numerical simulations of event-size distributions, there are however some recent gyrokinetic numerical simulations of the spectrum properties of heat flux~\cite{a41}-~\cite{a42}. However, the prediction of the full spectrum behavior of heat flux is out of the scope of the present work and will be addressed in a future publication.
 
In summary, this paper presents the first calculation of PDF tails of momentum flux and heat flux, which were shown to be significantly enhanced over the Gaussian prediction. This suggests that considerable transport is mediated by rare events of high amplitude. This is because even if the PDF tails have a low amplitude (rare events), these rare events are of high amplitude, possibly carrying significant transport. Since the PDF tails are enhanced over the Gaussian prediction in our case, these events are more likely to mediate considerable transport. The main point of the present work is to investigate the influence of different parameters on the coefficient ($\xi$). The results should be interpreted that for certain parameters, large scale events are more likely to be the main cause for transport while less important for other parameters. In all cases with parameters relevant for a tokamak plasma, the (enhanced) non-Gaussian PDF tail is of great importance. The overall amplitude is shown to be larger in ITG than in HM turbulence for reversed modon speed ($U<0$). 

\section{Acknowledgment}
The authors are indebted to J. Douglas for proofreading the manuscript. This research was supported by the Engineering and Physical Sciences Research Council (EPSRC) EP/D064317/1.

\end{document}